\pdfoutput=1

\newif\ifusenix
\newif\ifieee
\newif\ifarxiv
\newif\ifblind

\ifusenix
  \ieeefalse
  \arxivfalse
  \newif\ifCLASSOPTIONcompsoc\CLASSOPTIONcompsocfalse
  \newif\ifCLASSINFOpdf\CLASSINFOpdftrue
\else\ifieee
  \usenixfalse
  \arxivfalse
\else
  \arxivtrue
  \usenixfalse
  \ieeefalse
  \blindfalse
  \newif\ifCLASSOPTIONcompsoc\CLASSOPTIONcompsocfalse
  \newif\ifCLASSINFOpdf\CLASSINFOpdftrue
\fi\fi

\usenixfalse
\ieeefalse
\arxivtrue

\def\whenblind#1#2{\ifblind#1\else#2\fi}

\ifusenix
\documentclass[letterpaper,twocolumn,10pt]{article}
\usepackage{usenix-2020-09}
\fi

\ifieee
\documentclass[conference,compsoc]{IEEEtran}
\usepackage{amsmath}
\fi
\ifCLASSOPTIONcompsoc
  \usepackage[nocompress]{cite}
\else\ifarxiv\else
  \usepackage{cite}
\fi\fi

\ifarxiv
\documentclass{amsart}
\usepackage{amsmath}
\usepackage{fullpage}
\fi

\usepackage{mainpoints}
\HighlightMainPointsfalse

\ifHighlightMainPoints
\usepackage{checklist}
\fi

\usepackage{soul}

\usepackage[T1]{fontenc}

\ifCLASSINFOpdf
\else
\fi

\usepackage{amsmath}
\usepackage{url}
\usepackage{gnuplot-lua-tikz}
\usepackage{amsfonts}
\usepackage{algorithm}
\usepackage{algpseudocode}
\usepackage{tabularx}
\usepackage{flushend}

\algnewcommand\algorithmicforeach{\textbf{for each}}
\algdef{S}[FOR]{ForEach}[1]{\algorithmicforeach\ #1\ \algorithmicdo}

\begin{document}

%

\ifarxiv
\title{Blind Spots: Automatically Detecting Ignored Program Inputs}
\else
\title{Blind Spots: Identifying Exploitable Program Inputs}
\fi



\ifblind
\author{}
\else\ifarxiv
\author[H. Brodin]{Henrik Brodin}
\address{Trail of Bits, Stockholm, Sweden}
\email{henrik.brodin@trailofbits.com}
\author[E. Sultanik]{Evan Sultanik}
\address{Trail of Bits, Philadelphia, USA}
\email{evan.sultanik@trailofbits.com}
\author[M. Surovi\v{c}]{Marek Surovi\v{c}}
\address{Trail of Bits, Brno, Czech Republic}
\email{marek.surovic@trailofbits.com}
\else
\author{\IEEEauthorblockN{Henrik Brodin}
\IEEEauthorblockA{\textit{Trail of Bits}\\
New York, USA\\
henrik.brodin@trailofbits.com}
\and
\IEEEauthorblockN{Marek Surovi\v{c}}
\IEEEauthorblockA{\textit{Trail of Bits}\\
New York, USA\\
marek.surovic@trailofbits.com}
\and
\IEEEauthorblockN{Evan Sultanik}
\IEEEauthorblockA{\textit{Trail of Bits}\\
New York, USA\\
evan.sultanik@trailofbits.com\\
}
}
\fi\fi

\def\redacted#1{[{\color{gray}#1}]}
\def\PolyTracker{\whenblind{\redacted{tool name redacted for blind review}}{PolyTracker}}
\def\PolyFile{\whenblind{\redacted{tool name redacted for blind review}}{PolyFile}}
\def\blindcite#1{\whenblind{\cite{#1blind}}{\cite{#1}}}



%


\maketitle

\def\todo#1#2{
{\color{red}\textbf{ToDo[#1]:}~\textit{#2}}
}

\begin{abstract}
A \emph{blind spot} is any input to a program that can be arbitrarily mutated without affecting the program's output.
Blind spots can be used for steganography or to embed malware payloads.
If blind spots overlap file format keywords, they indicate parsing bugs that can lead to exploitable differentials.
For example, one could craft a document that renders one way in one viewer and a completely different way in another viewer.
They have also been used to circumvent code signing in Android binaries, to coerce certificate authorities to misbehave, and to execute HTTP request smuggling and parameter pollution attacks.
This paper formalizes the operational semantics of blind spots, leading to a technique based on dynamic information flow tracking that automatically detects blind spots.
An efficient implementation is introduced and evaluated against a corpus of over a thousand diverse PDFs parsed through M$\mu$PDF\footnote{\url{https://mupdf.com/}}, revealing exploitable bugs in the parser.
All of the blind spot classifications are confirmed to be correct and the missed detection rate is no higher than 11\%.
On average, at least 5\% of each PDF file is completely ignored by the parser.
Our results show promise that this technique is an efficient automated means to detect exploitable parser bugs, over-permissiveness and differentials.
Nothing in the technique is tied to PDF in general, so it can be immediately applied to other notoriously difficult-to-parse formats like ELF, X.509, and XML.
\end{abstract}


\ifieee\IEEEpeerreviewmaketitle\fi

\def\viz{\textit{vi}{\fontfamily{cmr}\textit{\&.}}}

\section{Introduction}

\MainPoint{We define a \emph{blind spot} as any input to a program that can be arbitrarily mutated without affecting the program's output.}
Blind spots are dangerous: they can be exploited for steganography and embedding malware payloads.
Steganographic attacks are notoriously difficult to detect automatically, but a brief manual analysis of five of the most popular archive file formats produced fifteen vulnerability disclosures enabled by steganography~\cite{vuksan10hiding}.
One such bug is the ``aCropalypse'' (CVE-2023-21036)~\cite{buchanan23exploiting}, publicly disclosed on March 18th, 2023.
It affects images created on both Google phones and Microsoft Windows, in which cropped image data persists in a blind spot and can subsequently be reconstructed.
The blind spot detection technique and tooling introduced in this paper has been used to detect both aCropalyptic images as well as vulnerable image generators~\cite{brodin23avoid}.

Blind spots can also be indicative of parser differentials, for instance, if two parsers exhibit \emph{different} blind spots for the same input.
Such differentials can be exploited by crafting a file that renders one way in one parser and a completely different way in another parser~\cite{albertini15abusing}.
For example, blind spots have been exploited to craft a PDF that can render one way in Adobe Acrobat but have different text when printed~\cite{sultanik16postscript}.
Blind spots have also been used to circumvent code signing in Android binaries~\cite{forristal14android}, to coerce certificate authorities to emit certificates for unauthorized Common Names~\cite{kaminsky10pki}, and to execute HTTP request smuggling and parameter pollution attacks~\cite{balduzzi11automated}.
Our results show that M$\mu$PDF, on average, ignores 5\% of each PDF file.
Some PDFs in our corpus had megabytes of ignored data that could be overwritten to store a malware payload.
Blind spots can also be useful: They can potentially be excluded as candidates for mutation when generating fuzz testing inputs,
similar to the Angora fuzzer's branch coverage maximization strategy~\cite{peng18angora}.


Blind spots are a generalization of the concept of \emph{file cavities} introduced by Albertini, \textit{et~al.}~\cite{albertini20abuse}: unused spaces in a file format that are created due to the structure of the surrounding data.
However, \MainPoint{unlike cavities, blind spots may be dependent on the program itself and its execution environment.}
For example, the image content of a JPEG file will be a blind spot to a parser that only reads its EXIF metadata.
Likewise, the EXIF metadata will be a blind spot to a program that converts JPEGs to another image format like BMP that
does not support embedded EXIF metadata.
A malware payload could be embedded within the EXIF data without affecting the image's rendering.
Similarly, a JPEG parser for which the EXIF data is \emph{not} a blind spot would likely render a specially crafted image \emph{differently} than a parser for which the EXIF data \emph{is} a blind spot.

Since they are associated with \emph{both} program input \emph{and} the program itself, \MainPoint{blind spots can
be indicators of parsing vulnerabilities.}
Parsers---particularly hand-generated ones---will often accept a superset of the grammar for which they were designed.
This manifests as a parser that accepts some inputs that are technically invalid according to the file format specification.
Sometimes this is intentional, in order to maximize compatibility with files generated by other, incorrectly implemented software, or to attempt to repair malformed documents.
For example, we discovered that an optimization in M$\mu$PDF will sometimes only
check the leading ``e'' in the \texttt{endobj} token;
it will eagerly accept \texttt{eXXXXX} and the PDF will still be parsed correctly,
despite the fact that the PDF standard prescribes the existence of the full token.
Such lexical permissiveness can lead to parser differentials and so-called \emph{file format
schizophrenia}~\cite{albertini15abusing}: when two implementations of a file format interpret the same input file differently.
A different PDF implementation will ignore the erroneous token and parse further to find the correct delimiter, resulting in different behavior and output.

This paper makes several contributions; it$\ldots$
\begin{enumerate}
  \item \MainPoint{formalizes the concept of program input blind spots};
  \item clarifies the difference between blind spots and file cavities: Blind spots are the set of input bytes
whose data flow never influences either control flow that leads to an output or an output itself for a given program;
  \item proposes a novel technique based on dynamic taint analysis to detect blind spots;
  \item demonstrates how this technique can be used to automatically detect lexical permissiveness and parser bugs;
  \item evaluates the approach, providing evidence that these blind spots are correct; and
  \item analyzes the properties of naturally occurring blind spots in PDFs parsed by M$\mu$PDF.
\end{enumerate}

\section{Background}

Dynamic Information Flow Tracking~(DIFT), also known as Dynamic Taint Analysis~(DTA), is a technique in which the flow of information through a program is modeled and tracked at runtime.
DIFT is a challenging problem; many modern approaches suffer from high implementation overhead, low accuracy, and/or low fidelity~\cite{brant21challenges}.
\emph{Universal taint analysis} is a form of DIFT that can track all input bytes throughout the execution of a program, mapping inputs to outputs~\cite{yang21drtaint}.

Our approach to detect blind spots relies on universal taint analysis: We want to identify the input bytes that affect neither program output nor any control flow that leads to a program output.
\MainPoint{There are significant challenges when performing universal taint analysis on real-world software.}
If a program accepts $n$~bytes of input, there are $O(2^n)$ ways that each of those bytes could combine to taint each program output.
Therefore, when the cyclomatic complexity of a program is large, the amount of combinations generated from even a small input is enormous.
For example, many document formats such as PDF use compression to keep file sizes small.
When tracking data flow through a PDF parser, there will be a significant number of such combinations as the data are decompressed.
This phenomenon is known as \emph{taint explosion}, which generally occurs when a function performs a large number of combinatorial operations on input data.

There are several tools and techniques for performing both DIFT and universal taint analysis.
Most tools are designed for fuzz testing use cases, which do not need to track many input bytes.
For example, the LLVM compiler framework has its own dataflow analysis instrumentation pass, the DataFlowSanitizer~(DFSan)~\cite{dfsan}, which is limited to tracking at most 8~inputs.
Even tools that were explicitly designed for universal taint analysis are unable to scale to input sizes sufficient for real-world inputs.
We have a further discussion of the limitations of current tooling in Section~\ref{sec:relatedwork}, below.

\MainPoint{This paper introduces a new technique that can efficiently achieve universal taint analysis for \emph{megabytes} of input.}

\section{Definitions and Formalization}
\label{sec:formalization}

This section formalizes the concept of input blind spots.
We do this with an extension to Schwartz, Avgerinos, and Brumley's \textsc{SimpIL} operational semantics for dynamic taint propagation~\cite{schwartz10all}.
Cheney, Ahmed, and Acar's semantics for \emph{dependency provenance}~\cite{cheney07provenance} could also be used to formalize blind spots, however,
the literature on provenance as dependency analysis is defined generically as to be compatible with a variety of use cases including databases, file systems, and scientific workflows.
\textsc{SimpIL}, on the other hand, more explicitly defines how program instrumentation would occur.
As such, it also directly informs the data structures necessary to achieve efficient universal taint analysis (see Section~\ref{sec:instrumentation}, below).
This is why we present the formalism using \textsc{SimpIL}'s operational semantics.

\subsection{An Extension to \textsc{SimpIL}}

\MainPoint{The original conception of dataflow analysis in \textsc{SimpIL} only tracks whether a given variable or
memory cell is tainted, not \emph{from whence} it is tainted.}
In order to detect blind spots, we need to additionally track exactly which input bytes influence output.
For example, consider the pseudocode in Algorithm~\ref{alg:controlflowtaints}.
The variable $a$ is tainted by program input on line~\ref{alg1:a}.
Moreover, the value of $a$ can indirectly cause hard-coded data ($d = 5$) to be written to output on
line~\ref{alg1:write} by virtue of the conditional on line~\ref{alg1:cond}.
Therefore, the first byte of the file \emph{cannot} be a blind spot, since its mutation \emph{can} affect output---despite the fact that the value of the first byte is never written to output.
Even if the \textsc{SimpIL} taint policy~(\textit{i.e.}, the rules by which taints are propagated) is sufficient to
detect that tainted inputs affected the output of the program, it is \emph{insufficient} to detect \emph{which}
inputs were responsible.
Therefore, we need to extend \textsc{SimpIL} to additionally track the provenance of a taint so that we can map a
complete data flow from inputs to outputs.

\begin{algorithm}
\caption{Tainted Control Flow}\label{alg:controlflowtaints}
\begin{algorithmic}[1]
\Procedure{TaintedControlFlow}{}
\State\label{alg1:a} $a \gets$ \Call{ReadInput}{1} \Comment{$a$ is tainted by the $1^{\mbox{\tiny st}}$ byte} 
\State $b \gets$ \Call{ReadInput}{1} \Comment{$b$ is tainted by the $2^{\mbox{\tiny nd}}$ byte} 
\State $c \gets a + b$
\If{$c \geq 42$}\label{alg1:cond}
\State $d \gets 5$
\State\label{alg1:write} \Call{WriteOutput}{$d$}
\EndIf
\EndProcedure
\end{algorithmic}
\end{algorithm}

\textsc{SimpIL} uses meta-syntactic variables to
represent an execution context. $\Delta$ is a mapping of variable
names to their values and $\mu$ is a mapping from memory addresses to their values. $\tau_\Delta$ and $\tau_\mu$ map
variable names and memory addresses to booleans~($\mathbf{T}|\mathbf{F}$) defining whether or not that value is tainted
in the current execution context.

\MainPoint{In order to track taint provenance, we use the concept of a \emph{taint label}\cite{peng18angora}: a unique identifier
for each instance of a tainted variable or memory address in an execution context.}
In addition to \emph{union} labels\cite{peng18angora} we also define \emph{canonical} labels.
A union taint is the result of the combination of two previously tainted values (\textit{e.g.},~the result of two
tainted variables being operands in a binary operation).
A canonical taint label is the result of a variable or memory address being assigned directly from a program input.

Let $\mathcal{I} = \{\langle s_0, i_0\rangle, \langle s_1, i_1\rangle, \ldots\}$ be the set of all possible program inputs, where each $s$ is the source (\textit{e.g.}, a file, pipe, socket, or environment variable) and $i$ is the offset within the source.
We extend the \textsc{SimpIL} notation with three new mappings to represent taint labels and track
provenance\footnote{The \textsc{SimpIL} semantics are already
replete with Greek letters. We have chosen $\varepsilon$ from the Greek word for
``labels''~($\epsilon\pi\iota\gamma\rho\alpha\phi\epsilon\varsigma$), 
  $\kappa$ from the word for ``canonical''~($\kappa\alpha\nu o\nu\iota\kappa o\varsigma$), 
  and $\gamma$
  from the word for ``parent''~($\gamma o\nu\epsilon\upsilon\varsigma$).}: 
\begin{enumerate}
  \item $\varepsilon: \Delta \cup \mu \rightarrow \mathbb{N}$ that maps variable names and memory addresses to unique
        taint labels\footnote{$\Delta$ and $\mu$ retain their original meanings from \textsc{SimpIL}.};
  \item $\kappa: \mathbb{N} \rightarrow \mathcal{I}$ that maps
  canonical taint labels to the information about their source; and
  \item $\gamma: \mathbb{N} \rightarrow \mathbb{N} \times \mathbb{N}$ that maps union taint labels to their parents.
\end{enumerate}
Note that the $\gamma$ mapping implicitly defines a directed acyclic graph~(DAG) where the out-degree of each vertex is at most two.
However, instructions that operate on tainted values could have an arity \emph{higher} than two, causing the result label to have more than two parents.
In such cases, we add multiple unions to the $\gamma$ mapping.
This is done for the sake of notational simplicity and does not affect our results.

\textsc{SimpIL} treats these mappings more like programmatic hashmaps than set theoretic functions.
As such, \textsc{SimpIL} uses the notation ``$\kappa[\ell]$'' for the value of taint label $\ell$ in mapping
$\kappa$.
For brevity, we shall continue this theme by using the notation $\ell \in \kappa$ to represent the fact that
$\ell$ is a key in the mapping $\kappa$, $\ell \notin \kappa$ to mean that $\ell$ is not a key in~$\kappa$, and
$|\kappa|$ to mean the number of key/value pairs in the mapping.

The zero taint label is reserved to represent untainted variables and memory.
An untainted variable $v$ will always lack source info and descend from the zero label:
\ifarxiv
\begin{displaymath}
  \tau_{\Delta}[v] = \begin{cases}
      \mathbf{F} & \varepsilon[v] \notin \kappa \wedge \gamma[\varepsilon[v]] = \langle 0, 0 \rangle,\\
      \mathbf{T} & \varepsilon[v] \in \kappa \vee (\gamma[\varepsilon[v]] = \langle i, j \rangle \wedge i + j > 0).
  \end{cases}
\end{displaymath}
\else
\begin{multline*}
  \tau_{\Delta}[v] = \begin{cases}
      \mathbf{F} & \varepsilon[v] \notin \kappa \wedge \gamma[\varepsilon[v]] = \langle 0, 0 \rangle,\\
      \mathbf{T} & \varepsilon[v] \in \kappa \vee (\gamma[\varepsilon[v]] = \langle i, j \rangle \wedge i + j > 0).
  \end{cases}
\end{multline*}
\fi

The \textsc{SimpIL} \emph{taint policy}~(see Table~III from~\cite{schwartz10all}) and semantics are modified
to update these mappings on every taint status change.
For example, the updated semantics for reading from input and for executing binary operations are given in
Figure~\ref{fig:semantics}.
\begin{figure*}
\begin{displaymath}
  \frac{
    \overbrace{v\ \text{is input from}\ src \in \mathcal{I}\quad
    \varepsilon^\prime = \varepsilon[v \leftarrow |\varepsilon| + 1]\quad
    \kappa^\prime = \kappa[\varepsilon^\prime[v] \leftarrow src]}^{\text{\scriptsize\textsc{SimpIL} notation for the
    computation performed by the \textsc{Input} operation}}
  }{
    \underbrace{\mu,\Delta,\varepsilon,\kappa,\gamma \leadsto \mu,\Delta,\varepsilon^\prime,\kappa^\prime,
    \gamma}_{\text{\parbox{2in}{\centering\scriptsize\textsc{SimpIL} notation for the\\updated execution state after the
  operation}}}\quad
    \underbrace{\mu,\Delta \vdash \text{get\_input}(src) \Downarrow
    v}_{\text{\parbox{2in}{\centering\scriptsize\textsc{SimpIL} notation for
    the evaluation of\\expression get\_input$(src)$ to value $v$ in context $\mu,\Delta$}}}
  }
  \ \textsc{Input}
\end{displaymath}
\medskip
\begin{displaymath}
  \frac{
    \mu,\Delta \vdash e_1 \Downarrow v_1\quad
    \mu,\Delta \vdash e_2 \Downarrow v_2\quad
    v^\prime=v_1\Diamond_b v_2\quad
    \varepsilon^\prime = \varepsilon[v^\prime \leftarrow |\varepsilon| + 1]\quad
    \gamma^\prime = \gamma\big[\varepsilon^\prime[v^\prime] \leftarrow \langle\varepsilon[v_1],
    \varepsilon[v_2]\rangle\big]
  }{
    \mu,\Delta,\varepsilon,\kappa,\gamma \leadsto \mu,\Delta,\varepsilon^\prime,\kappa,\gamma^\prime\quad
    \mu,\Delta \vdash e_1 \Diamond_b e_2 \Downarrow v^\prime
  }
  \ \text{\textsc{Binop}}
\end{displaymath}
  \caption{\textsc{SimpIL} operational semantics for reading input and executing binary operators (see Figure~1
  of~\cite{schwartz10all}) updated to
  include data flow provenance tracking.
  When a value $v$ is read from an input source $src$, we create a new,
    unique canonical taint label for $v$ and set its taint source info in $\kappa$ to $src$. When a binary operator
    $\Diamond_b$ is applied to expressions $e_1 = v_1$ and
    $e_2 = v_2$ resulting in the value $v^\prime$, we create a new, unique union taint label for $v^\prime$ and set its
    parents to be the taint labels associated with values $v_1$ and $v_2$.}
  \label{fig:semantics}
\end{figure*}

\subsection{Mapping Taint Sources to Sinks}

These mappings allow us to track the entire provenance of a taint in any execution context.
As defined above, the $\gamma$ mapping implicitly creates a DAG of labels, representing the dataflow through the program:
The program inputs that affect a tainted variable or memory address are its taint label's topmost ancestors in the $\gamma$~DAG.
A recursive function $\psi: \mathbb{N} \rightarrow 2^{\mathcal{I}}$ can map taint
labels to all of their ancestral sources:
\begin{displaymath}
  \psi[\ell] = \begin{cases}
                 \{\kappa[\ell]\} & \ell \in \kappa,\\
                 \bigcup_{p \in \gamma[\ell]} \psi[p] & \gamma[\ell] \neq \langle 0, 0 \rangle,\\
                 \emptyset & \text{otherwise.}
  \end{cases}
\end{displaymath}

\MainPoint{When we observe that the program writes to output, we use the $\psi$ mapping to record which
inputs, if any, tainted the output.}
This can be accomplished by enumerating the canonical ancestors of the output labels by traversing the $\gamma$~DAG.
This allows us to construct a complete mapping of taint sources to sinks.
Note that any taint sources without associated sinks can be arbitrarily mutated without affecting the output.

\subsection{A Definition of Program Input Blind Spots}

Let $\Omega$ be the set of taint labels written to output during execution.
Then a blind spot is the set of all potential program inputs that did not affect the output:

\begin{equation}\begin{aligned}\label{equ:blindspot}
  \mathcal{I}_{\text{\textsc{$\neg$Read}}} &= \{\iota \in \mathcal{I} : (\forall \ell \in \kappa : \kappa[\ell] \neq \iota)\} \\
  \mathcal{I}_{\text{\textsc{$\neg$InOutput}}} &= \{ \kappa[\ell] : \ell \in \kappa
     \wedge (\forall
     \ell^\prime \in \Omega : \kappa[\ell] \notin \psi[\ell^\prime])\} \\
  \mathcal{I}_{\text{\textsc{BlindSpot}}} &= \mathcal{I}_{\text{\textsc{$\neg$Read}}} \cup \mathcal{I}_{\text{\textsc{$\neg$InOutput}}}.
\end{aligned}\end{equation}

As we mentioned above, \textsc{SimpIL} includes a \emph{taint policy} specifying the rules by which taints are
propagated.
\MainPoint{The taint policy will affect our definition of blind spots.}
For example, let us again consider the pseudocode in Algorithm~\ref{alg:controlflowtaints}.
The variable $a$ is tainted by program input on line~\ref{alg1:a}, which we can now specify in our \textsc{SimpIL}
extension as $\kappa[\varepsilon[a]] \neq \emptyset$.
Recall that the value of $a$ can indirectly cause hard-coded data ($d = 5$) to be written to output on line~\ref{alg1:write}.
Therefore, $a$ cannot be a blind spot, since its mutation can affect output.
However, the semantics by which the taint policy propagates taint through conditionals will affect whether our
extension of \textsc{SimpIL} will consider the output to be tainted by $a$, because the value of $a$ itself is never
written to output.

We resolve this discrepancy by enforcing the following constraint on blind spot taint policies: The
taint labels of every variable and memory address that affect the program's control-flow will be unioned with all
labels created in the branch they influence.
In other words, in addition to the definition of blind spots in Equation~\eqref{equ:blindspot}, \MainPoint{a blind spot cannot
influence control flow that leads to a program output}.
Updated operational semantics for the conditional operator that implement this policy are given in Figure~\ref{fig:conditionals}.

\begin{figure*}
\begin{displaymath}
  \frac{
    \overbrace{\mu,\Delta \vdash e \Downarrow v}^{\text{\ifarxiv\begin{minipage}{0.2\hsize}\centering\fi the conditional expression $e$ evaluates to $v$\ifarxiv\end{minipage}\fi}}\quad
    \overbrace{\varepsilon^\prime = \varepsilon[u \leftarrow |\varepsilon| + \varepsilon[u] : \forall u \in \varepsilon]}^{\text{\ifarxiv\begin{minipage}{0.2\hsize}\centering\fi create a new taint label for every existing label\ifarxiv\end{minipage}\fi}}\quad
    \overbrace{\gamma^\prime = \gamma\big[\varepsilon^\prime[u] \leftarrow \langle\varepsilon[u], \varepsilon[v]\rangle : \forall u \in \varepsilon\big]}^{\text{union every existing label with the taints of $v$}}
  }{
    \mu,\Delta,\varepsilon,\kappa,\gamma \leadsto \mu,\Delta,\varepsilon^\prime,\kappa,\gamma^\prime
  }
  \ \text{\textsc{PreCond}}
\end{displaymath}
  \caption{Updated \textsc{SimpIL} operational semantics to enforce
  the taint policy that every input that affects control-flow will be
  unioned with all labels created in the branch they
  influence. The \textsc{PreCond} rule is executed before every conditional rule (\textsc{TCond} and \textsc{FCond} in Figure~1 of~\cite{schwartz10all}).}
  \label{fig:conditionals}
\end{figure*}

A trace of Algorithm~\ref{alg:controlflowtaints} showing the iterative updates to the execution context is given in
Table~\ref{tab:executionexample}.
It demonstrates how the blind spot taint policy propagates taints from variables in the path
condition---variables that have affected control flow leading to the current state (\textit{e.g.},~variable $c$
on line~\ref{alg1:cond})---to variables that would otherwise not be tainted (\textit{e.g.},~variable~$d$).
This reduces false-positive blind spot classifications, since it captures tainted variables that indirectly cause
output.

\def\stackedwrapper#1{\begin{minipage}[m]{2cm}\centering#1\end{minipage}}
\def\stackedsettwo#1#2{\stackedwrapper{
\begin{tikzpicture}
  \node[anchor=center] (a) at (0,0) {$#1$\rlap{,}};
  \node[anchor=north] (b) at (a.base) {$#2$};
  \node[anchor=base east,xshift=0.5em] at (a.base west) {$\{$};
  \node[anchor=base west,xshift=-0.5em] at (b.base east) {$\}$};
\end{tikzpicture}}}
\def\stackedsetthree#1#2#3{\stackedwrapper{
\begin{tikzpicture}
  \node[anchor=center] (c) at (0,0) {$#2$\rlap{,}};
  \node[anchor=base] (a) at (c.north) {$#1$\rlap{,}};
  \node[anchor=north] (b) at (c.base) {$#3$};
  \node[anchor=base east,xshift=0.5em] at (a.base west) {$\{$};
  \node[anchor=base west,xshift=-0.5em] at (b.base east) {$\}$};
\end{tikzpicture}}}
\def\stackedsetfour#1#2#3#4{\stackedwrapper{
\begin{tikzpicture}
  \node[anchor=center] (c) at (0,0) {$#2$\rlap{,}};
  \node[anchor=base] (a) at (c.north) {$#1$\rlap{,}};
  \node[anchor=north] (b) at (c.base) {$#3$\rlap{,}};
  \node[anchor=north] (d) at (b.base) {$#4$};
  \node[anchor=base east,xshift=0.5em] at (a.base west) {$\{$};
  \node[anchor=base west,xshift=-0.5em] at (d.base east) {$\}$};
\end{tikzpicture}}}

\begin{table*}
  \begingroup\ifarxiv\scriptsize\else\footnotesize\fi
  \def\firstbyte{\langle1^{\text{\tiny st}}\ \text{byte of input}\rangle}
  \def\secondbyte{\langle2^{\text{\tiny nd}}\ \text{byte of input}\rangle}
  \begin{center}\begin{tabular}{c||l|c|c|c|c|c@{}}
    Line & \textsc{Statement} & $\Delta$ & $\tau_\Delta$ & $\varepsilon$ & $\kappa$ & $\gamma$ \\ \hline
    \hline
    1 & start & $\{\}$ & $\{\}$ & $\{\}$ & $\{\}$ & $\{\}$ \\ \hline

    2 &
    $a \gets$ \textsc{ReadInput}$(1)$ &
    $\{a \rightarrow 40\}$ & $\{a \rightarrow \mathbf{T}\}$ &
    $\{a \rightarrow 1\}$ & $\{1~\rightarrow~\firstbyte\}$ & $\{\}$ \\ \hline

    3 & $b \gets$ \textsc{ReadInput}$(1)$ &
    \stackedsettwo{a \rightarrow 40}{b \rightarrow 12} &
    \stackedsettwo{a \rightarrow \mathbf{T}}{b \rightarrow \mathbf{T}} &
    \stackedsettwo{a \rightarrow 1}{b \rightarrow 2} &
    \hspace*{-4em}\stackedsettwo{1 \rightarrow \firstbyte}{2 \rightarrow \secondbyte} &
    $\{\}$ \\ \hline

    4 & $c \gets a + b$ &
    \stackedsetthree{a \rightarrow 40}{b \rightarrow 12}{c \rightarrow 52} &
    \stackedsetthree{a \rightarrow \mathbf{T}}{b \rightarrow \mathbf{T}}{c \rightarrow \mathbf{T}} &
    \stackedsetthree{a \rightarrow 1}{b \rightarrow 2}{c \rightarrow 3} &
    \hspace*{-4em}\stackedsettwo{1 \rightarrow \firstbyte}{2 \rightarrow \secondbyte} &
    $\{3 \rightarrow \langle 1, 2 \rangle\}$
    \\\hline

    5 & $\mathbf{if}\ c \geq 42\ \mathbf{then}$ &
    \stackedsetthree{a \rightarrow 40}{b \rightarrow 12}{c \rightarrow 52} &
    \stackedsetthree{a \rightarrow \mathbf{T}}{b \rightarrow \mathbf{T}}{c \rightarrow \mathbf{T}} &
    \stackedsetthree{a \rightarrow 1}{b \rightarrow 2}{c \rightarrow 3} &
    \hspace*{-4em}\stackedsettwo{1 \rightarrow \firstbyte}{2 \rightarrow \secondbyte} &
    $\{3 \rightarrow \langle 1, 2 \rangle\}$
    \\\hline

    6 & $d \gets 5$ &
    \stackedsetfour{a \rightarrow 40}{b \rightarrow 12}{c \rightarrow 52}{d \rightarrow 5} &
    \stackedsetfour{a \rightarrow \mathbf{T}}{b \rightarrow \mathbf{T}}{c \rightarrow \mathbf{T}}{d \rightarrow
    \mathbf{T}} &
    \stackedsetfour{a \rightarrow 1}{b \rightarrow 2}{c \rightarrow 3}{d \rightarrow 4} &
    \hspace*{-4em}\stackedsettwo{1 \rightarrow \firstbyte}{2 \rightarrow \secondbyte} &
    \stackedsettwo{3 \rightarrow \langle 1, 2 \rangle}{4 \rightarrow \langle 3, 0 \rangle}
  \end{tabular}\end{center}\endgroup
  \caption{Execution context trace for Algorithm~\ref{alg:controlflowtaints}. Note on line~6 that, despite being
  assigned a constant value of~5, the $d$ variable (taint label~4) is in fact tainted by variable $c$ (taint label~3).
  This is because the path condition to line~6 depends on $c$ from the conditional branch on line~5.
  Therefore, neither of the first two bytes of input are blind spots.}
  \label{tab:executionexample}
\end{table*}

\section{Implementation}
\label{sec:instrumentation}

Thus far we have developed formal semantics for blind spots and discovered some necessary taint propagation policies
to detect them.
The next step is to automatically instrument a parser to extract the data flow information necessary to classify
input byte regions as blind spots.
\MainPoint{We gather this data flow information by performing universal taint analysis.}

\MainPoint{\PolyTracker\ifblind\else~\cite{sultanik19two}\fi\ is an LLVM-based dynamic analysis tool that we have developed for extracting ground truth information from programs.}
It is open-source and available at \whenblind{\redacted{git repository URL redacted for blind review}}{\url{https://github.com/trailofbits/polytracker}}.

\MainPoint{\PolyTracker\ automatically adds instrumentation to a program such that, when the program is executed, it
produces runtime artifacts that can be analyzed to track the data flows of all input bytes.}
It is an extension of the LLVM DataFlowSanitizer~(DFSan)~\cite{dfsan}, a generalized dynamic data flow
analysis instrumentation tool.
\PolyTracker\ has previously been used to label the semantic purpose of functions in a parser~\blindcite{harmon20toward}.

Originally, DFSan supported tracking at most $2^{16}$ taint labels at a time.
This limit was only sufficient to track at most several hundred input bytes at once.
Over the course of 2021, DFSan underwent a significant refactor in order to make its memory layout compatible with other LLVM sanitizers~\cite{dfsan-refactor};
this refactor reduced the effective number of taints it could track to $2^3 = 8$.
This restriction was acceptable for DFSan's primary use case at the time: data flow analysis for fuzz testing, but \emph{not} for universal taint analysis and detecting blind spots.

We forked \PolyTracker\ off of the final, pre-refactor version of DFSan.
For the remainder of this section, our discussion of DFSan will refer to this version.
This version of DFSan works by creating a region of ``shadow'' memory that can store a taint label associated with every address on the stack and heap.
For every instruction in the program, DFSan checks its operands to see if they have associated taint labels in shadow memory.
If the labels are different, it means that the operands were tainted by different input data flows, and DFSan will create a new label that represents the union of the two.
DFSan also has mechanisms for propagating taint information across function calls (\textit{e.g.}, by appending taint labels as function arguments), as well as models for taint propagation through uninstrumented system calls.

DFSan uses a $2^{16}\times 2^{16}$ matrix to store the unions generated when instructions mix taint labels.
Element $i, j$ of the matrix holds the value of the label produced by the union of labels $i$ and $j$.
This matrix representation is computationally efficient but becomes prohibitively large as the maximum number of taint labels grows.
For $n$-bit taint labels, the matrix will require $$\Theta\left(2^n \times 2^n \times \frac{n}{8}\right) = \Theta(2^{2n - 3}n)$$ bytes of memory.
This is the reason for DFSan's limit of $2^{16}$ taint labels: Increasing the limit to $2^{32}$ labels would require over 73~\emph{exabytes} of RAM to store the union matrix.
\MainPoint{We need a way to increase this $2^{16}$ limit by at least a few orders of magnitude.}

Our solution to this problem arises from the $\gamma$ mapping we added to the \textsc{SimpIL} operational semantics.
We create a memory-mapped file where each taint label has a fixed-size entry containing the indexes of its parent labels.
\PolyTracker\ adds additional instrumentation to:
\begin{itemize}
  \item tag input sources as canonical taints (building the $\kappa$ mapping from our semantics);
  \item track the taint labels that are written to output; and
  \item track which taint labels affect control flow.
\end{itemize}

The algorithm for determining blind spots from a program trace is given in Algorithm~\ref{alg:blindspots}.
Set $L$ denotes the set of all taint labels in a program trace. Consequently set $B$ denotes all taint labels that did not affect control flow.
The algorithm iterates through labels in $L$ in descending order. On line $5$, $\ell \notin B$ means that taint label $\ell$ or its descendant affected control flow, while $\ell \in \Omega$ means that taint label $\ell$ was written to output. If either of these is true $\ell$ and its parents are removed from $B$ on line $6$. Finally on line $9$, the set all blind spots $\mathcal{I}_{\text{\textsc{BlindSpot}}}$ is the set of all inputs which have their associated taint label in $B$.
The algorithm runs in $O(n)$ time where $n$ is the number of taint labels created during the trace.
This has proven sufficient to detect blind spots in all programs and inputs on which we have experimented.

\begin{algorithm}
\caption{Enumerate Blind Spots}\label{alg:blindspots}
\begin{algorithmic}[1]
\Ensure{$\mathcal{I}_{\text{\textsc{BlindSpot}}}$ is the set of blind spots in the trace}
\Procedure{BlindSpots}{$\Omega$, $\varepsilon$, $\kappa$, $\gamma$}
\State $L \gets \{\varepsilon[v]: v \in \varepsilon\} \cup \{\ell: \ell \in \kappa\}$\Comment{the set of all taint labels in the trace}
\State $B \gets \{\ell \in L: \neg \text{\textsc{AffectedControlFlow}}(\ell)\}$\Comment{the set of all labels that did not affect control flow}
\ForEach{$\ell \in \text{\textsc{SortDescending}}(L)$}
  \If{$\ell \notin B$ $\vee$ $\ell \in \Omega$}
    \Comment{$\ell$ cannot be a blind spot because it or one of its descendants affected output}
    \State $B \gets B \setminus (\{\ell\} \cup \gamma[\ell])$
  \EndIf
\EndFor
\State $\mathcal{I}_{\text{\textsc{BlindSpot}}} \gets \{\kappa[\ell] : \ell \in B\}$
\EndProcedure
\end{algorithmic}
\end{algorithm}

\subsection{Related Work}
\label{sec:relatedwork}

There are several existing projects that achieve universal taint tracking, using various methods.
Two of the best maintained and easiest to use are AUTOGRAM~\cite{autogram2016} and TaintGrind~\cite{taintgrind}.
However, the former is limited to analysis within the Java virtual machine and the latter suffers from unacceptable runtime overhead when tracking as few as several bytes at a time.
For example, we ran \texttt{mutool}, a utility in the M$\mu$PDF project, using TaintGrind over a corpus of medium sized PDFs, and in every case the tool had to be halted after over twenty-four hours of execution for operations that would normally complete in milliseconds without instrumentation.

There are also existing tools for performing dynamic program analysis
via QEMU~\cite{bellard05qemu}, such as PANDA~\cite{dolangavitt15panda}
and DECAF(++)~\cite{davanian19decaf}, both of which have taint tracking extensions.
However, being an emulation framework rather than virtualization, QEMU incurs a runtime overhead of about~15\% just to execute a binary, not including any instrumentation~\cite{ahmed20measuring}.
After adding the program instrumentation necessary to enable fuzz testing, QEMU was observed to have over three times the runtime overhead of equivalent compile-time instrumentation~\cite{nagy21breaking}.

Symbolic execution engines like Triton~\cite{saudel15triton} and SymCC~\cite{poeplau2020symcc} have also been used for data flow analysis.
Symbolic execution could be extended to detect blind spots, \textit{e.g.}, by making all input bytes symbolic and observing all data that is written.
The input bytes associated with any symbolic data that is either written or included in the path condition during a write \emph{is not} a blind spot.

DRTaint~\cite{yang21drtaint} is a recently published tool that can also perform universal taint tracking.
It adds a minimal amount of runtime instrumentation to create runtime artifacts that can be post-processed to extract any data flow.
The authors do not quantify the exact overhead of DRTaint, but Figure~5 from their paper suggests at
least a 60x slowdown compared to the uninstrumented program.
It is also unclear whether this instrumentation was sufficient to reconstruct all data flows.
This is consistent with earlier techniques such as Dytan~\cite{clause07dytan} that reported a 50x slowdown when tracking as few as 64~taint labels.

\section{Evaluation}
\label{sec:analysis}

The previous sections introduced a method for identifying the blind spots of a program input.
How accurate is this blind spot classifier?
\MainPoint{Since we do not have pre-labeled ground truth for the blind spots of an input,
we need to develop statistical estimates for the confusion matrix of our classifier.}

\MainPoint{We focus our evaluation on the PDF file format, for several reasons.}
\begin{enumerate}
  \item The PDF file format is old and complex, has had many revisions, and enjoys numerous independent implementations.
  This has led to differentials that necessitate lexical permissiveness for interoperability~\cite{wyatt21demystifying}.
  \item PDF is a container format that allows embedding of other formats like JPEG, providing more opportunity for blind spots.
  \item The GovDocs corpus~\cite{garfinkel09bringing} provides thousands of real-world PDFs generated by a diversity of software.
\end{enumerate}

\MainPoint{We instrumented \texttt{mutool}, a utility in the popular M$\mu$PDF project, using \PolyTracker\ to detect blind spots.}
Next, we ran the instrumented utility on 1,087 PDFs sampled from the GovDocs corpus to render the PDFs to PostScript.
The PDFs totaled over 622~MB and averaged 572~KB each.
The largest file was 9.6~MB.
We discovered a total of 33.5~MB of blind spots, averaging 30.8~KB per file, some files having zero, and
one file having 1.07~MB of blind spots.

\MainPoint{PostScript, a vector image format, was chosen rather than a raster format like JPEG
or PNG because the relative lack of compression would help prevent explosion of taint union labels and thereby reduce runtime.}
Runtime of the instrumented program was less than one minute for each PDF.
We would expect to get similar results when rendering to JPEG or PNG, however, since blind spots are dependent on
execution, there could be some discrepancies.
For example, since fonts can be embedded in both PDF and PostScript but cannot be embedded in JPEG or PNG, one might
expect to see more blind spots related to fonts if one were to have rendered to a different file format.

\subsection{Classification Error}

\MainPoint{We validate blind spots by iteratively mutating each classified blind spot byte in the input file and
re-running the original, uninstrumented program again.} See Figure~\ref{fig:validationexample} for a notional example of this mutation process.
If the PostScript output produced from the mutated input is different from the output of the unmodified file, then our classified blind
spot is incorrect (Type~I error), since any mutation inside a blind spot should, by definition, not affect program output.

\MainPoint{We also sample bytes from \emph{outside} of our classified blind spots and mutate them, similarly.}
If a byte outside a blind spot can be arbitrarily mutated, it is likely a missed detection (Type~II error).
However, it is not tractable to mutate and verify all possible combinations of input bytes, since this would amount to testing the power set of all bytes, running in $\Theta(2^n)$ time.
While detecting blind spots was relatively quick---several hours to process the 1,087~PDF corpus---, validating the blind spots by mutating the inputs required about a month of computation.

It might be the case that a byte outside a blind spot is in fact a byte that can be \emph{almost}
arbitrarily mutated, but has some undetected data dependency on another byte.
For example, a source code comment is this form of input to a compiler, since the bytes within the comment can be arbitrarily changed without affecting the behavior of the compiler \emph{as long as} the bytes do not contain the comment delimiter. 
Therefore our reported Type~I error rate is a tight bound on the actual Type~I error, but our Type~II error rate is a loose upper bound on the true Type~II error.

\begin{figure}
  \begingroup
  \setul{0.5ex}{0.3ex}
  \def\mutated#1{\begingroup
    \setulcolor{red}\ul{#1}
  \endgroup}
  \begin{center}
    \parbox{\ifarxiv 0.375\hsize\else 0.75\hsize\fi}{\small\tt
      L\mutated{o}r\mutated{e}m ip\mutated{s}um\mutated{~}dolo\mutated{r} sit a\mutated{m}et,
      conse\mutated{c}t\mutated{e}tur
      adipiscing elit\mutated{,} \mutated{se}d
      d\mutated{o} e\mutated{ius}mo\mutated{d}
      te\mutated{m}por\mutated{~}%
      in\mutated{ci}didu\mutated{nt} ut
    \mutated{l}ab\mutated{o}re \mutated{et}
    d\mutated{olo}re
      ma\mutated{gna} a\mutated{l}i\mutated{qu}a.
      \tikz[remember picture]{\node[coordinate,yshift=0.4\baselineskip] (bsstart) {};}\mutated{Ut enim ad minim
      veniam,}
      \tikz[remember picture]{\node[coordinate,yshift=-0.1\baselineskip] (bsend) {};}
      q\mutated{u}is \mutated{no}st\mutated{ru}d\mutated{~e}xer\mutated{c}it\mutated{at}ion\mutated{~}
      u\mutated{lla}mc\mutated{o~l}abo\mutated{r}is
      n\mutated{i}si ut \mutated{ali}quip e\mutated{x~e}a
      com\mutated{mo}do conseq\mutated{uat.} D\mutated{ui}s au\mutated{te~i}rure dolor in
      repreh\mutated{e}nd\mutated{e}rit in vol\mutated{upt}a\mutated{te~}
      \mutated{v}el\mutated{i}t e\mutated{s}se
      cil\mutated{lu}m do\mutated{lor}e eu \mutated{f}ugiat \mutated{n}ulla \mutated{par}iat\mutated{ur.}
    }
    \begin{tikzpicture}[remember picture,overlay]
      \node[draw=blue,line width=2pt,fit=(bsstart) (bsend)] (outline) {};
      \node[yshift=-3em,fill=blue,text=white] at (outline.south) (caption) {Detected Blind
      Spot};
      \draw[draw=blue,line width=2pt,->] (caption.north) -- (outline.south);
    \end{tikzpicture}
  \end{center}
  \caption{Notional example of validating detected blind spots.
  Underlined bytes are iteratively mutated and re-parsed.
  Bytes outside of blind spots are randomly selected for mutation, but all bytes within a blind spot are mutated. If
  a byte \emph{inside} a blind spot is mutated and produces a different result, then that blind spot classification
  was a false-positive (Type~I error). If a byte \emph{outside} a blind spot can be arbitrarily mutated, then it is a missed
  detection (Type~II error).}
  \label{fig:validationexample}
  \endgroup
\end{figure}

We mutated all 33.5~million blind spot bytes classified in the corpus, and \MainPoint{all mutated blind spots
produced identical output to the original file for a 0\% false-positive rate.} Of the bytes \emph{not}
classified as blind spots, 89\% did affect the output.
Therefore, the false-negative rate is bounded above by 11\%.
A significant number of these missed detections are likely data that can be mutated \emph{almost} arbitrarily, but have some data dependency that can affect output that our random mutations did not exercise.

\subsection{Blind Spot Content}

What is the content of PDF blind spots for M$\mu$PDF?

In the GovDocs PDF corpus parsed by M$\mu$PDF, we detect 63,194 unique blind spot prefixes of length at most seven bytes, and 338,943 unique byte
sequences of length at most seven that precede blind spots.

Consider the bytes that occur at the start of a blind spot; the most common of these are listed in Table~\ref{tab:prefixes}.
They are all one byte long, meaning that there is a diversity of content at the start of blind spots.
The most common byte sequences that \emph{precede} a blind spot, listed in Table~\ref{tab:suffixes}, are more interesting:
most are multi-byte, and they comprise many PDF tokens like \texttt{endobj}.
\MainPoint{This means that bytes following certain tokens are often or always ignored by the parser.}

\begin{table}
\begin{center}\ifarxiv\footnotesize\else\scriptsize\fi
  \begin{tabular}{c|c|c}
    \textsc{Bytes} & \# Blind Spots & Total Frequency \\ \hline\hline
    \texttt{\textbackslash r} & 561451 & 6745095 \\ \hline
    \texttt{\textbackslash n} & 52175 & 4996436 \\ \hline
    \texttt{\textbackslash x20} & 47752 & 17967270 \\ \hline
    \texttt{\textbackslash x00} & 9504 & 7658663 \\ \hline
    \texttt{\textbackslash x11} & 5232 & 3325230 \\ \hline
    \texttt{\textbackslash x08} & 3930 & 3611559 \\ \hline
    \texttt{\textbackslash x06} & 3629 & 3104109 \\ \hline
    \texttt{\textbackslash x09} & 3572 & 2981341 \\ \hline
    \texttt{\%} & 3145 & 3041586 \\ \hline
    \texttt{\textbackslash x12} & 2932 & 2847588 \\ \hline
    \texttt{\#} & 2859 & 3399908 \\ \hline
    \texttt{\textbackslash x05} & 2772 & 2873564 \\ \hline
    \texttt{\textbackslash x02} & 2640 & 3007837 \\ \hline
    \texttt{\textbackslash x0F} & 2607 & 2944161 \\ \hline
    \texttt{\textbackslash x14} & 2594 & 3132850 \\ \hline
    \texttt{\textbackslash xF0} & 2351 & 2930162 \\ \hline
    \texttt{a} & 2348 & 4959732 \\ \hline
    \texttt{!} & 2271 & 3172472 \\ \hline
    \texttt{\textbackslash xA3} & 2108 & 2783202 \\ \hline
    \texttt{4} & 2074 & 4848767 \\ \hline
    \texttt{\textbackslash x13} & 2024 & 2788459 \\ \hline
  \end{tabular}
\end{center}
\caption{The twenty most common blind spot prefixes of length at most seven bytes.}
\label{tab:prefixes}
\end{table}

\begin{table}
\begin{center}\ifarxiv\footnotesize\else\scriptsize\fi
  \begin{tabular}{c|c|c}
    \textsc{Bytes} & \# Blind Spots & Total Frequency \\ \hline\hline
    \texttt{n} & 300057 & 5811224 \\ \hline
    \texttt{\textbackslash\ n} & 299785 & 629959 \\ \hline
    \texttt{00000\textbackslash\ n} & 299621 & 555279 \\ \hline
    \texttt{f} & 236173 & 3738217 \\ \hline
    \texttt{\textbackslash\ f} & 235736 & 552315 \\ \hline
    \texttt{65535\textbackslash\ f} & 205564 & 470855 \\ \hline
    \texttt{m} & 66860 & 4242668 \\ \hline
    \texttt{dstream} & 66588 & 189282 \\ \hline
    \texttt{00001\textbackslash\ f} & 26343 & 44612 \\ \hline
    \texttt{\textbackslash r} & 12839 & 6745095 \\ \hline
    \texttt{endobj\textbackslash r} & 11001 & 527199 \\ \hline
    \texttt{e} & 7056 & 7495303 \\ \hline
    \texttt{be} & 6419 & 41673 \\ \hline
    \texttt{Adobe} & 6418 & 11545 \\ \hline
    \texttt{\textbackslash x00\textbackslash x0EAdobe} & 6417 & 9142 \\ \hline
    \texttt{\textbackslash x02} & 5533 & 3007837 \\ \hline
    \texttt{\textbackslash x08} & 4185 & 3611559 \\ \hline
    \texttt{\textbackslash x00\textbackslash x02} & 3787 & 82034 \\ \hline
    \texttt{00000\textbackslash\ f} & 3770 & 6838 \\ \hline
    \texttt{\textbackslash x01\textbackslash x00\textbackslash x02} & 3707 & 29366 \\ \hline
  \end{tabular}
\end{center}
\caption{The twenty most common byte sequences of length at most seven preceding a blind spot.}
\label{tab:suffixes}
\end{table}

Now let us consider the unique suffix/prefix pairs that occur adjacent to the start of a blind spot.
There are 1,029,129 unique pairs of these byte sequences.
If we sort them by frequency, the pair
\begin{displaymath}
  \overbrace{\ldots\text{\texttt{e}}}^{\text{\scriptsize\llap{not a blind spot}}}
  \hspace*{-1pt}\underbrace{\text{\texttt{ndobj}}\ldots}_{\text{\scriptsize\rlap{blind spot}}}
\end{displaymath}
is in the top 0.01\% of such pairs.
``\texttt{endobj}'' is a PDF token used to delimit the end of an object in the document model.
The fact that this token is split across a blind spot boundary so frequently is indicative of, at best, intentional
lexical permissiveness on the part of the parser, and, at worst, a bug.
Other interesting blind spot contexts within the top hundredth of the first percentile include the entire
\texttt{endobj} token, if preceded by a carriage return.
Any whitespace after the \texttt{stream} token is ignored.
The \texttt{obj} token is completely ignored if preceded by a space.
Similarly, the PDF dictionary delimiters $<<$ and $>>$ are frequently skipped, \textit{e.g.}, at the beginning of a PDF object.
\MainPoint{This simple contextual blind spot frequency analysis can discover parsing errors and differentials.}

\subsection{Blind Spot Context}

How frequent are blind spots, and where do they occur in PDFs?

Figure~\ref{fig:BlindSpotsByFileSize} plots the number of blind spots in the PDF corpus as a function of file size.
\MainPoint{This suggests that the number of blind spot bytes in a typical PDF is constant.}
Note, however, that blind spots in PDFs can be arbitrarily large, since the PDF format permits the inclusion of
arbitrary binary blobs that do not have to be connected to the document object model~(DOM)~\cite{wyatt21demystifying}.

\begin{figure*}
\begin{center}
\input{graphs/cavities_by_filesize}
\end{center}
\caption{Proportion of PDF bytes parsed by M$\mu$PDF that are in blind spots as a function of file size.
This suggests that the number of blind spot bytes in a typical M$\mu$PDF-parsed PDF is constant.}
\label{fig:BlindSpotsByFileSize}
\end{figure*}

Figure~\ref{fig:BlindSpotsBySize} plots a histogram of the contiguous size of blind spot regions in the corpus.
\MainPoint{The majority of blind spots are small, but a nontrivial number of blind spots are over 1~KB.}
The average blind spot is 42~bytes long with a standard deviation of 1.72~KB.
This demonstrates the fact that PDF is a container format that can contain arbitrarily large binary blobs that do not have to contribute to PDF rendering.

\begin{figure*}
\begin{center}
\input{graphs/cavity_size}
\end{center}
\caption{Histogram of the sizes of contiguous PDF blind spots for M$\mu$PDF. The majority of blind spots are single bytes, but
a nontrivial number of blind spots are over 1~kilobyte. The average blind spot is 42~bytes long with a standard
deviation of 1.72~kilobytes.
This demonstrates the fact that PDF is a container format that can contain arbitrarily large binary blobs that do not have to contribute to PDF rendering.}
\label{fig:BlindSpotsBySize}
\end{figure*}

Figure~\ref{fig:BlindSpotsByPosition} is a histogram of the normalized position of blind spot bytes in their files:
the blind spot's byte offset divided by the file size.
\MainPoint{In our experiments with the M$\mu$PDF renderer translating to PostScript, the majority of PDF blind spots
are at the beginning and ends of the files.}
This is not surprising since the beginning of a PDF typically includes metadata that is not necessary for rendering,
and the end of the PDF typically includes an XREF table that can be ignored, particularly if the PDF is not malformed.
Anecdotally, we have observed in the GovDocs corpus that PDF generators often add additional metadata objects that do not contribute to rendering, typically toward the end of the file.

\begin{figure*}
\begin{center}
\input{graphs/cavity_position}
\end{center}
\caption{Normalized position of blind spots in PDF files. With the M$\mu$PDF renderer, blind spots are most frequent
at the beginning and end of PDF files.
The large number of blind spots toward the end of files can be explained by their context within the PDF cross-reference~(XREF) table (\textit{q.v.}~Table~\ref{tab:derivations}), which is not strictly necessary for rendering.}
\label{fig:BlindSpotsByPosition}
\end{figure*}

Figure~\ref{fig:BlindSpotSizeByPosition} combines the two previous figures by comparing the mean contiguous blind
spot size to the normalized position in the PDF.
Despite the most blind spot bytes being at the beginning and ends of the PDFs, \MainPoint{the \emph{longest} blind
spots tend to be in the first 10--20\% of the file, but not immediately at the beginning.}
The abundance of blind spots toward the end of the document tend to be small.
In order to explain them, we need to look for patterns in their semantic context.

\begin{figure*}
\begin{center}
\begin{tikzpicture}[gnuplot]
\path (0.000,0.000) rectangle (16.510,6.096);
\gpcolor{color=gp lt color border}
\gpsetlinetype{gp lt border}
\gpsetdashtype{gp dt solid}
\gpsetlinewidth{1.00}
\draw[gp path] (1.320,0.985)--(1.500,0.985);
\draw[gp path] (15.774,0.985)--(15.594,0.985);
\node[gp node right] at (1.136,0.985) {$0$};
\draw[gp path] (1.320,1.671)--(1.500,1.671);
\draw[gp path] (15.774,1.671)--(15.594,1.671);
\node[gp node right] at (1.136,1.671) {$20$};
\draw[gp path] (1.320,2.357)--(1.500,2.357);
\draw[gp path] (15.774,2.357)--(15.594,2.357);
\node[gp node right] at (1.136,2.357) {$40$};
\draw[gp path] (1.320,3.043)--(1.500,3.043);
\draw[gp path] (15.774,3.043)--(15.594,3.043);
\node[gp node right] at (1.136,3.043) {$60$};
\draw[gp path] (1.320,3.729)--(1.500,3.729);
\draw[gp path] (15.774,3.729)--(15.594,3.729);
\node[gp node right] at (1.136,3.729) {$80$};
\draw[gp path] (1.320,4.415)--(1.500,4.415);
\draw[gp path] (15.774,4.415)--(15.594,4.415);
\node[gp node right] at (1.136,4.415) {$100$};
\draw[gp path] (1.320,5.101)--(1.500,5.101);
\draw[gp path] (15.774,5.101)--(15.594,5.101);
\node[gp node right] at (1.136,5.101) {$120$};
\draw[gp path] (1.320,5.787)--(1.500,5.787);
\draw[gp path] (15.774,5.787)--(15.594,5.787);
\node[gp node right] at (1.136,5.787) {$140$};
\draw[gp path] (1.320,0.985)--(1.320,1.165);
\draw[gp path] (1.320,5.787)--(1.320,5.607);
\node[gp node center] at (1.320,0.677) {0\%};
\draw[gp path] (2.765,0.985)--(2.765,1.165);
\draw[gp path] (2.765,5.787)--(2.765,5.607);
\node[gp node center] at (2.765,0.677) {10\%};
\draw[gp path] (4.211,0.985)--(4.211,1.165);
\draw[gp path] (4.211,5.787)--(4.211,5.607);
\node[gp node center] at (4.211,0.677) {20\%};
\draw[gp path] (5.656,0.985)--(5.656,1.165);
\draw[gp path] (5.656,5.787)--(5.656,5.607);
\node[gp node center] at (5.656,0.677) {30\%};
\draw[gp path] (7.102,0.985)--(7.102,1.165);
\draw[gp path] (7.102,5.787)--(7.102,5.607);
\node[gp node center] at (7.102,0.677) {40\%};
\draw[gp path] (8.547,0.985)--(8.547,1.165);
\draw[gp path] (8.547,5.787)--(8.547,5.607);
\node[gp node center] at (8.547,0.677) {50\%};
\draw[gp path] (9.992,0.985)--(9.992,1.165);
\draw[gp path] (9.992,5.787)--(9.992,5.607);
\node[gp node center] at (9.992,0.677) {60\%};
\draw[gp path] (11.438,0.985)--(11.438,1.165);
\draw[gp path] (11.438,5.787)--(11.438,5.607);
\node[gp node center] at (11.438,0.677) {70\%};
\draw[gp path] (12.883,0.985)--(12.883,1.165);
\draw[gp path] (12.883,5.787)--(12.883,5.607);
\node[gp node center] at (12.883,0.677) {80\%};
\draw[gp path] (14.329,0.985)--(14.329,1.165);
\draw[gp path] (14.329,5.787)--(14.329,5.607);
\node[gp node center] at (14.329,0.677) {90\%};
\draw[gp path] (15.774,0.985)--(15.774,1.165);
\draw[gp path] (15.774,5.787)--(15.774,5.607);
\node[gp node center] at (15.774,0.677) {100\%};
\draw[gp path] (1.320,5.787)--(1.320,0.985)--(15.774,0.985)--(15.774,5.787)--cycle;
\node[gp node center,rotate=-270] at (0.292,3.386) {Mean Contiguous Blind Spot Size (Bytes)};
\node[gp node center] at (8.547,0.215) {Normalized Blind Spot Start Position (Byte Offset $\div$ File Size)};
\gpcolor{rgb color={0.580,0.000,0.827}}
\draw[gp path] (1.320,0.985)--(1.320,2.628)--(2.765,2.628)--(2.765,0.985)--cycle;
\draw[gp path] (2.765,0.985)--(2.765,4.828)--(4.211,4.828)--(4.211,0.985)--cycle;
\draw[gp path] (4.211,0.985)--(4.211,3.294)--(5.656,3.294)--(5.656,0.985)--cycle;
\draw[gp path] (5.656,0.985)--(5.656,2.734)--(7.102,2.734)--(7.102,0.985)--cycle;
\draw[gp path] (7.102,0.985)--(7.102,3.828)--(8.547,3.828)--(8.547,0.985)--cycle;
\draw[gp path] (8.547,0.985)--(8.547,3.623)--(9.992,3.623)--(9.992,0.985)--cycle;
\draw[gp path] (9.992,0.985)--(9.992,3.759)--(11.438,3.759)--(11.438,0.985)--cycle;
\draw[gp path] (11.438,0.985)--(11.438,3.292)--(12.883,3.292)--(12.883,0.985)--cycle;
\draw[gp path] (12.883,0.985)--(12.883,2.298)--(14.329,2.298)--(14.329,0.985)--cycle;
\draw[gp path] (14.329,0.985)--(14.329,1.418)--(15.774,1.418)--(15.774,0.985)--cycle;
\draw[gp path] (2.043,2.400)--(2.043,2.856);
\draw[gp path] (1.953,2.400)--(2.133,2.400);
\draw[gp path] (1.953,2.856)--(2.133,2.856);
\draw[gp path] (3.488,3.898)--(3.488,5.759);
\draw[gp path] (3.398,3.898)--(3.578,3.898);
\draw[gp path] (3.398,5.759)--(3.578,5.759);
\draw[gp path] (4.934,2.915)--(4.934,3.673);
\draw[gp path] (4.844,2.915)--(5.024,2.915);
\draw[gp path] (4.844,3.673)--(5.024,3.673);
\draw[gp path] (6.379,2.474)--(6.379,2.994);
\draw[gp path] (6.289,2.474)--(6.469,2.474);
\draw[gp path] (6.289,2.994)--(6.469,2.994);
\draw[gp path] (7.824,3.318)--(7.824,4.339);
\draw[gp path] (7.734,3.318)--(7.914,3.318);
\draw[gp path] (7.734,4.339)--(7.914,4.339);
\draw[gp path] (9.270,3.169)--(9.270,4.077);
\draw[gp path] (9.180,3.169)--(9.360,3.169);
\draw[gp path] (9.180,4.077)--(9.360,4.077);
\draw[gp path] (10.715,3.398)--(10.715,4.121);
\draw[gp path] (10.625,3.398)--(10.805,3.398);
\draw[gp path] (10.625,4.121)--(10.805,4.121);
\draw[gp path] (12.161,3.027)--(12.161,3.556);
\draw[gp path] (12.071,3.027)--(12.251,3.027);
\draw[gp path] (12.071,3.556)--(12.251,3.556);
\draw[gp path] (13.606,2.156)--(13.606,2.440);
\draw[gp path] (13.516,2.156)--(13.696,2.156);
\draw[gp path] (13.516,2.440)--(13.696,2.440);
\draw[gp path] (15.051,1.390)--(15.051,1.447);
\draw[gp path] (14.961,1.390)--(15.141,1.390);
\draw[gp path] (14.961,1.447)--(15.141,1.447);
\gpcolor{color=gp lt color border}
\node[gp node right] at (14.306,5.453) {Average Blind Spot Length};
\gpcolor{rgb color={0.000,0.620,0.451}}
\draw[gp path] (14.490,5.453)--(15.406,5.453);
\draw[gp path] (1.320,2.400)--(1.466,2.400)--(1.612,2.400)--(1.758,2.400)--(1.904,2.400)%
  --(2.050,2.400)--(2.196,2.400)--(2.342,2.400)--(2.488,2.400)--(2.634,2.400)--(2.780,2.400)%
  --(2.926,2.400)--(3.072,2.400)--(3.218,2.400)--(3.364,2.400)--(3.510,2.400)--(3.656,2.400)%
  --(3.802,2.400)--(3.948,2.400)--(4.094,2.400)--(4.240,2.400)--(4.386,2.400)--(4.532,2.400)%
  --(4.678,2.400)--(4.824,2.400)--(4.970,2.400)--(5.116,2.400)--(5.262,2.400)--(5.408,2.400)%
  --(5.554,2.400)--(5.700,2.400)--(5.846,2.400)--(5.992,2.400)--(6.138,2.400)--(6.284,2.400)%
  --(6.430,2.400)--(6.576,2.400)--(6.722,2.400)--(6.868,2.400)--(7.014,2.400)--(7.160,2.400)%
  --(7.306,2.400)--(7.452,2.400)--(7.598,2.400)--(7.744,2.400)--(7.890,2.400)--(8.036,2.400)%
  --(8.182,2.400)--(8.328,2.400)--(8.474,2.400)--(8.620,2.400)--(8.766,2.400)--(8.912,2.400)%
  --(9.058,2.400)--(9.204,2.400)--(9.350,2.400)--(9.496,2.400)--(9.642,2.400)--(9.788,2.400)%
  --(9.934,2.400)--(10.080,2.400)--(10.226,2.400)--(10.372,2.400)--(10.518,2.400)--(10.664,2.400)%
  --(10.810,2.400)--(10.956,2.400)--(11.102,2.400)--(11.248,2.400)--(11.394,2.400)--(11.540,2.400)%
  --(11.686,2.400)--(11.832,2.400)--(11.978,2.400)--(12.124,2.400)--(12.270,2.400)--(12.416,2.400)%
  --(12.562,2.400)--(12.708,2.400)--(12.854,2.400)--(13.000,2.400)--(13.146,2.400)--(13.292,2.400)%
  --(13.438,2.400)--(13.584,2.400)--(13.730,2.400)--(13.876,2.400)--(14.022,2.400)--(14.168,2.400)%
  --(14.314,2.400)--(14.460,2.400)--(14.606,2.400)--(14.752,2.400)--(14.898,2.400)--(15.044,2.400)%
  --(15.190,2.400)--(15.336,2.400)--(15.482,2.400)--(15.628,2.400)--(15.774,2.400);
\gpcolor{color=gp lt color border}
\draw[gp path] (1.320,5.787)--(1.320,0.985)--(15.774,0.985)--(15.774,5.787)--cycle;
\gpdefrectangularnode{gp plot 1}{\pgfpoint{1.320cm}{0.985cm}}{\pgfpoint{15.774cm}{5.787cm}}
\end{tikzpicture}
\end{center}
\caption{Contiguous blind spot size as a function of its position in the PDF file parsed by M$\mu$PDF. Error bars correspond to
the standard deviation of blind spot sizes in that portion of the file. The longest blind spots tend to be in the second tenth of the file.}
\label{fig:BlindSpotSizeByPosition}
\end{figure*}

In order to assign a semantic context to each byte of input, we generate a parse tree for each PDF using \PolyTracker's sister tool,
\PolyFile\footnote{Open-source and available at \whenblind{\redacted{git repository URL redacted for blind review}}{\url{https://github.com/trailofbits/polyfile}}.}~\blindcite{sultanik19two}.
\MainPoint{Each byte in the input PDF corresponds to one or more \emph{parse tree derivations}: unique paths through the
PDF parse tree.}
A byte could have more than one derivation, for instance, if the input file is a \emph{polyglot}---a file that is
valid in two or more formats~\cite{albertini15abusing}.
PDFs are particularly easy to turn into polyglots, and many legitimate PDF generators exploit this fact.
For example, it is common to produce valid PDFs that are \emph{also} valid ZIP archives that, when extracted, contain additional files related to the document.
Therefore, a byte might have one derivation in the PDF parse tree and have a different but completely valid derivation in the ZIP parse tree.
An example of a parse tree derivation is given in Figure~\ref{fig:derivation}.

\begin{figure}
\begin{center}\scriptsize
  \framebox{application/pdf} $\rightarrow$ \framebox{PDFObject}
  $\rightarrow$ \framebox{PDFDictionary} $\rightarrow$ \framebox {KeyValuePair} $\rightarrow$ \framebox{Value}
  $\rightarrow$ \framebox{PDFList} $\rightarrow$ \framebox{PDFList} $\rightarrow$ \framebox{PSInt}.
\end{center}
  \caption{M$\mu$PDF's parse tree derivation of a byte representing an integer in a list of lists that is a value in a PDF
  dictionary in a PDF object.}
  \label{fig:derivation}
\end{figure}

For each unique parse tree derivation, we count the number of blind spot bytes that occur in that derivation.
\MainPoint{The most frequent derivation containing blind spots is {\scriptsize\framebox{application/pdf}}, the root of
the PDF parse tree.}
This means that the majority of PDF blind spots occur in portions of the file that have no semantic purpose.
Blind spot locality might be explained by the fact that PDF parsers are resilient to both leading and trailing garbage bytes before and after the PDF file.
Also, as we saw above, the majority of naturally occurring blind spot bytes are at the beginning and end of the file.

The frequency of every unique parse tree derivation is presented in Figure~\ref{fig:SemanticDistribution}.
Blind spots overwhelmingly occur in a small number of derivations.
\MainPoint{However, the long tail demonstrates that blind spots can and do occur in many diverse derivations.}

\begin{figure*}
\begin{center}
\input{graphs/semantic_distribution}
\end{center}
\caption{Each point is a \emph{parse tree derivation}---a unique path through a M$\mu$PDF PDF parse tree---whose $y$-axis value is the number of times a blind spot occurred in that derivation. Blind spots overwhelmingly occur in a small number of derivations, yet there is a long tail demonstrating that blind spots can and do occur in many diverse derivations.}
\label{fig:SemanticDistribution}
\end{figure*}

The most frequent parse tree derivations for PDF blind spots are given in Table~\ref{tab:derivations}.
\MainPoint{Almost all of the derivations descend from the PDFObject node.}
This is unsurprising, since PDF objects can contain streams of arbitrary binary data.
PDF objects also do not need to be connected to the root of the PDF document object model, nor do they need to be used in any way for rendering.
The third most common context in which blind spots occur is within the cross-reference~(XREF) table.
The XREF table is used by the parser to quickly look up the file offsets of PDF objects, decreasing load times.
However, the XREF table is not strictly necessary to parse a PDF, and almost all parsers are resilient to errors or omissions in the XREF table.
Therefore, it is unsurprising that there would be many blind spots within the XREF table.
The XREF table usually occurs toward the end of the PDF, explaining the positional distribution in Figure~\ref{fig:BlindSpotsByPosition}.

The second most frequent derivation for blind spots are in PDF dictionaries.
PDF dictionaries can contain arbitrary key/value pairs which are often used for metadata that is not necessary for rendering (\textit{e.g.},~timestamps).
Dictionaries can and often do contain redundant information.
For example, the length of a PDF object stream can either be specified as a key/value pair in the preceding object dictionary or implicitly defined by the location of a required termination token.
If both are specified, then they must agree.
However, if a length is specified in the dictionary which \emph{does not} agree with the position of the termination token, then most parsers will ignore the specified length and defer to the token position~\cite{wolf11omg}, making the dictionary entry a blind spot.

\begin{table*}
  \begin{tabularx}{\hsize}{>{\scriptsize}X|>{\scriptsize}r}
    \textsc{Derivation} & \textsc{\# Bytes} \\ \hline\hline
    application/pdf & 559430 \\\hline
    application/pdf $\rightarrow$ PDFObject $\rightarrow$ PDFDictionary $\rightarrow$ KeyValuePair $\rightarrow$ Key & 346492 \\\hline
    application/pdf $\rightarrow$ XRefTable & 303713 \\\hline
    application/pdf $\rightarrow$ PDFObject $\rightarrow$ PSBytes $\rightarrow$ image/jpeg & 120125 \\\hline
    application/pdf $\rightarrow$ PDFObject $\rightarrow$ PDFDictionary $\rightarrow$ KeyValuePair $\rightarrow$ Value $\rightarrow$ PDFObjRef & 112948 \\\hline
    application/pdf $\rightarrow$ PDFObject $\rightarrow$ PDFDictionary $\rightarrow$ KeyValuePair $\rightarrow$ Value $\rightarrow$ PDFLiteral & 101912 \\\hline
    application/pdf $\rightarrow$ PDFObject $\rightarrow$ PDFDictionary $\rightarrow$ KeyValuePair $\rightarrow$ Value $\rightarrow$ PSInt & 56427 \\\hline
    application/pdf $\rightarrow$ PDFObject $\rightarrow$ PDFDictionary $\rightarrow$ KeyValuePair $\rightarrow$ Value $\rightarrow$ PDFList $\rightarrow$ PDFObjRef & 54860 \\\hline
    application/pdf $\rightarrow$ PDFObject $\rightarrow$ PDFDictionary $\rightarrow$ KeyValuePair $\rightarrow$ Value $\rightarrow$ PDFList $\rightarrow$ PSInt & 49158 \\\hline
    application/pdf $\rightarrow$ PDFObject $\rightarrow$ PDFList $\rightarrow$ PDFObjRef & 37978 \\\hline
    application/pdf $\rightarrow$ PDFObject $\rightarrow$ PDFDictionary $\rightarrow$ KeyValuePair $\rightarrow$ Value $\rightarrow$ PDFList $\rightarrow$ PDFLiteral & 30841 \\\hline
    application/pdf $\rightarrow$ PDFObject $\rightarrow$ PDFDictionary $\rightarrow$ KeyValuePair $\rightarrow$ Value $\rightarrow$ PDFDictionary $\rightarrow$ KeyValuePair $\rightarrow$ Key & 26643 \\\hline
    application/pdf $\rightarrow$ PDFObject $\rightarrow$ FlateDecode $\rightarrow$ DecodedStream $\rightarrow$ PSBytes $\rightarrow$ application/zlib & 25803 \\\hline
    application/pdf $\rightarrow$ PDFObject $\rightarrow$ PDFDictionary $\rightarrow$ KeyValuePair $\rightarrow$ Value $\rightarrow$ PSFloat & 17474 \\\hline
    application/pdf $\rightarrow$ PDFObject $\rightarrow$ PDFDictionary $\rightarrow$ KeyValuePair $\rightarrow$ Value $\rightarrow$ PDFList $\rightarrow$ PDFList $\rightarrow$ PSInt & 14489 \\\hline
    application/pdf $\rightarrow$ PDFObject $\rightarrow$ PDFDictionary $\rightarrow$ KeyValuePair $\rightarrow$ Value $\rightarrow$ PSBytes $\rightarrow$ text/plain & 12653 \\\hline
    application/pdf $\rightarrow$ PDFObject $\rightarrow$ FlateDecode & 11446 \\\hline
    application/pdf $\rightarrow$ PDFObject $\rightarrow$ PDFDictionary $\rightarrow$ KeyValuePair $\rightarrow$ Value $\rightarrow$ PDFList $\rightarrow$ PSFloat & 10991 \\\hline
    application/pdf $\rightarrow$ PDFObject $\rightarrow$ PDFDictionary $\rightarrow$ KeyValuePair $\rightarrow$ Value $\rightarrow$ PDFDictionary $\rightarrow$ KeyValuePair $\rightarrow$ Value $\rightarrow$ PDFLiteral & 10052 \\\hline
    application/pdf $\rightarrow$ PDFObject $\rightarrow$ PDFDictionary $\rightarrow$ KeyValuePair $\rightarrow$ Value $\rightarrow$ PDFList $\rightarrow$ PDFDictionary $\rightarrow$ KeyValuePair $\rightarrow$ Key & 8250 \\\hline
    application/pdf $\rightarrow$ PDFObject $\rightarrow$ PDFDictionary $\rightarrow$ KeyValuePair $\rightarrow$ Value $\rightarrow$ PDFDictionary $\rightarrow$ KeyValuePair $\rightarrow$ Value $\rightarrow$ PSFloat & 7743 \\\hline
    application/pdf $\rightarrow$ PDFObject $\rightarrow$ PSBytes $\rightarrow$ application/octet-stream & 7328 \\\hline
    application/pdf $\rightarrow$ PDFObject $\rightarrow$ PSBytes $\rightarrow$ image/jp2 & 6147 \\\hline
    application/pdf $\rightarrow$ PDFObject $\rightarrow$ PSBytes $\rightarrow$ text/plain & 5748 \\\hline
    application/pdf $\rightarrow$ PDFObject $\rightarrow$ PDFDeciphered $\rightarrow$ image/jpeg & 4900 \\\hline
    application/pdf $\rightarrow$ PDFObject $\rightarrow$ PDFDictionary $\rightarrow$ KeyValuePair $\rightarrow$ Value $\rightarrow$ PDFList $\rightarrow$ PSBytes $\rightarrow$ text/plain & 4845 \\\hline
    application/pdf $\rightarrow$ PDFObject $\rightarrow$ PDFDictionary $\rightarrow$ KeyValuePair $\rightarrow$ Value $\rightarrow$ PDFDictionary $\rightarrow$ KeyValuePair $\rightarrow$ Value $\rightarrow$ PSInt & 4722 \\\hline
    application/pdf $\rightarrow$ PDFObject $\rightarrow$ PDFDictionary $\rightarrow$ KeyValuePair $\rightarrow$ Value $\rightarrow$ PDFList $\rightarrow$ PDFDictionary $\rightarrow$ KeyValuePair $\rightarrow$ Value $\rightarrow$ PDFLiteral & 3518 \\\hline
    application/pdf $\rightarrow$ PDFObject $\rightarrow$ PDFDictionary $\rightarrow$ KeyValuePair $\rightarrow$ Value $\rightarrow$ PDFDictionary $\rightarrow$ KeyValuePair $\rightarrow$ Value $\rightarrow$ PDFList $\rightarrow$ PSInt & 3439 \\\hline
    application/pdf $\rightarrow$ PDFObject $\rightarrow$ PDFDeciphered $\rightarrow$ text/plain & 3361 \\\hline
    application/pdf $\rightarrow$ PDFObject $\rightarrow$ PDFDictionary $\rightarrow$ KeyValuePair $\rightarrow$ Value $\rightarrow$ PDFList $\rightarrow$ PDFList $\rightarrow$ PDFObjRef & 2936 \\\hline
    application/pdf $\rightarrow$ PDFObject $\rightarrow$ PSBytes $\rightarrow$ image/x-portable-bitmap & 2806
  \end{tabularx}
  \caption{The PDF parse tree derivations for M$\mu$PDF containing the most blind spot bytes. These primarily descend from PDFObject.
    Almost all of the derivations descend from the PDFObject node.
    This is unsurprising, since PDF objects can contain streams of arbitrary binary data.
    PDF objects also do not need to be connected to the root of the PDF document object model, nor do they need to be used in any way for rendering.
  }
  \label{tab:derivations}
\end{table*}

\section{Conclusions}

This paper defined the concept of blind spots: inputs to a program that can be arbitrarily mutated without affecting the program's output.
Operational semantics for blind spots were formalized by extending \textsc{SimpIL}~\cite{schwartz10all}.
An efficient implementation capable of automatically detecting blind spots, \PolyTracker, was introduced.
It works by adding instrumentation for performing dynamic information flow tracking~(DIFT) to a program.

The technique was evaluated by detecting blind spots in the popular M$\mu$PDF parser over a corpus of over a thousand diverse PDFs~\cite{garfinkel09bringing}.
There were zero false-positive blind spot classifications, and the missed detection rate was bounded above by 11\%.
On average, at least 5\% of each PDF file was completely ignored by the parser; blind spots that could be repurposed for steganography or embedding malware payloads.

Future work includes extending the approach to detect inputs that can \emph{almost} arbitrarily be mutated without affecting output, like source code comments.
Using the revealed blinds spots to identify parser differentials, by comparing the blind spots of different instrumented parsers, would provide insight into vulnerabilities stemming from parsers interpreting the same file differently.
The current implementation injects its DIFT instrumentation at the LLVM/IR level.
Therefore, it is limited to programs that can be compiled using LLVM, or binaries that can be lifted to LLVM/IR.
It would be useful to apply the technique to runtime instrumentation that could be applied to a black-box binary.

Our results show promise that this technique could be an efficient automated means to detect parser bugs and differentials.
Nothing in the technique is tied to PDF in general, so it can be immediately applied to other notoriously difficult-to-parse formats like ELF, X.509, and XML.

\ifCLASSOPTIONcompsoc
  \section*{Acknowledgments}
\else
  \section*{Acknowledgment}
\fi

\ifblind
\redacted{Acknowledgements redacted for blind review.}
\else
This research was supported in part by the Defense Advanced Research Projects Agency~(DARPA) SafeDocs program as a subcontractor to Galois under HR0011-19-C-0073.
Many thanks to Michael Brown, Trent Brunson, Filipe Casal, Peter Goodman, Kelly Kaoudis, Lisa Overall, Stefan Nagy, Bill Harris, Nichole Schimanski, Mark Tullsen, Walt Woods, Peter Wyatt, Ange Albertini,
and Sergey Bratus for their invaluable feedback on the approach and tooling.
Thanks to Ange Albertini for suggesting \textit{\guillemotleft angles morts\guillemotright}---French for ``blind spots''---to name the concept.
Special thanks to Carson Harmon, the original creator of \PolyTracker, whose ideas and discussions germinated this
research.
\fi


\ifieee\bibliographystyle{IEEEtran}\else\ifarxiv\bibliographystyle{amsplain}\else\bibliographystyle{plain}\fi\fi
\bibliography{cavities.bib}

\end{document}